\documentstyle[tighten,floats,aps,epsf,preprint]{revtex}

\begin{document}
\title
{Structure of excited states of $^{11}$Be studied with Antisymmetrized 
Molecular Dynamics}

\author{Y. Kanada-En'yo}

\address{Institute of Particle and Nuclear Studies, \\
High Energy Accelerator Research Organization,\\
1-1 Oho, Tsukuba, Ibaraki 305-0801, Japan}

\author{H. Horiuchi}

\address{Department of Physics, Kyoto University, Kyoto 606-01, 
Japan}

\maketitle
\begin{abstract}
The structures of the ground and  excited states of $^{11}$Be were
studied with a microscopic method of antisymmetrized molecular dynamics.
The theoretical results reproduce the abnormal parity of the ground state
and predict various kinds of excited states.
We suggest a new negative-parity band with a well-developed clustering
structure which reaches high-spin states.
Focusing on a $2\alpha$ clustering structure, we investigated structure
 of the ground and excited states. We point out that molecular orbits 
play important roles for the intruder ground state 
and the low-lying $2\hbar \omega$ states.
The features of the breaking of $\alpha$ clusters were also studied with 
the help of data for Gamow-Teller transitions.
\end{abstract}


\section{Introduction}
Recently, information on the excited states of light unstable nuclei 
has increased rapidly \cite{KORSHENINNIKOV,AOI,FUJIWARA,FREER,MILIN}.
In the excited states of light unstable nuclei,
an exotic molecular structure of light unstable nuclei is one of 
the attractive subjects in experimental and theoretical research.
For example, molecular structure 
has been suggested to appear in neutron-rich nuclei, such as $^{10}$Be and 
$^{12}$Be \cite{KORSHENINNIKOV,FREER,MILIN,OERTZEN,OERTZENa,DOTE,ENYOf,ENYOg,ITAGAKI,OGAWA}. 
W. Von Oertzen et al. \cite{OERTZEN,OERTZENa} proposed  
a kind of exotic clustering structure with a $2\hbar\omega$ configuration 
in the excited states of $^{11}$Be.
However, there have been few microscopic studies on the excited states of 
$^{11}$Be.

Needless to say, $\alpha$-cluster cores are very important in 
the ground and excited states of $^{11}$Be as well as in $^9$Be and $^{10}$Be. 
However, we should not forget the breaking of $\alpha$-clusters in 
$^{11}$Be because there are many valence neutrons around the 
$\alpha$-clusters. 
The experimental data concerning the $\beta$-decay strength are 
very useful to estimate the 
breaking of the 2$\alpha$ clustering structure in Be isotopes,
because Gamow-Teller transitions from Li to Be are not allowed if 
the $\alpha$-cluster cores in the daughter state of Be are completely 
ideal ones with a simple $(0s)^4$ configuration.
It is very interesting that the recently measured $\beta$-decay strength
from $^{11}$Li to $^{11}$Be indicates 
significant breaking of $\alpha$-clusters in the excited 
states of $^{11}$Be \cite{AOI}. 
Therefore, it is important to also study the
breaking of $\alpha$-clusters due to the 
surrounding neutrons as well as the development of clustering in $^{11}$Be.

We should point out another interesting feature in $^{11}$Be.
Abnormal parity of the ground state in $^{11}$Be
has been known for a long time.
Namely, the spin parity of the ground state is $1/2^+$, which 
seems to be inconsistent with the ordinary shell-model picture
in which $^{11}$Be with 7 neutrons may have a $1/2^-$ state as the ground
state.
It has long been a problem why parity inversion occurs in $^{11}$Be.
As possible reasons for this parity inversion,  such effects as the 
halo structure, a core deformation, clustering and paring are suggested.
For example, the energy gain of a 
positive parity state of $^{11}$Be is discussed in Refs 
\cite{SAGAWA,SAGAWAb,FUKU,DOTEb,RAGNARSSON}. 
However, the ground-state structure has not been sufficiently studied by
fully microscopic calculations without such assumptions as the existence of 
core nuclei.
Although the clustering structure must be important in Be isotopes, 
it is difficult for mean-field approaches to describe developed 
clustering structures.
With a theoretical method of antisymmetrized molecular dynamics,
Dot\'{e} et al. have studied the abnormal parity ground state of $^{11}$Be
without assuming cores or the stability of the mean field. 
However, in their work\cite{DOTE}, 
the effects of angular-momentum projections and three-body forces were 
approximately estimated  by perturbative treatments.

Our aim is to make a systematic research on the structure of the ground and 
excited states of $^{11}$Be based on the microscopic 
calculations.
An important point is that the theoretical approach should be free from 
such model assumptions as stability of the mean field, 
and the existence of inert cores or clusters.
First of all, traditional mean-field approaches are not useful
to study the developed clustering structure in Be isotopes. 
With cluster models, the clustering structure of the 
excited states of $^9$Be and $^{10}$Be has been successfully explained
\cite{ITAGAKI,OGAWA,SEYA,OKABE,ARAI} 
by assuming 2-$\alpha$ cores and surrounding neutrons.
Since the assumption of 2 $\alpha$-cluster cores is not appropriate
to discuss the breaking of $\alpha$-cores, 
these cluster models are not sufficient for investigations of $^{11}$Be 
with many valence neutrons.
In fact, it is difficult to use them to  directly calculate the 
experimental data of the $\beta$-decay strength from $^{11}$Li. 
With these models it may not be possible to describe
the recently discovered excited state at 8.04 MeV with the strong 
$\beta$ transition strength from $^{11}$Li.

We have applied a theoretical approach of 
antisymmetrized molecular dynamics (AMD). 
The AMD method has already proved to be  useful 
for studying the structures of light nuclei 
\cite{DOTE,ENYOg,ENYOa,ENYObc,ENYOe,ENYOsup}.
Within this framework, we do not need such model assumptions as
inert cores, clusters, nor axial symmetries, because the basis 
wave functions of a nuclear system are written by Slater determinants
where all centers of the Gaussian-type spatial part of single-particle 
wave functions are free parameters. 
In AMD studies of neutron-rich nuclei, 
we investigated the structures of Be isotopes 
\cite{DOTE,ENYObc,ENYOsup}.
In AMD calculations, many kinds of experimental data for 
nuclear structure have been reproduced. 
Due to the flexibility of the AMD wave function,
we have succeeded to describe the structure changes 
between shell-model-like states and clustering states 
with an increase in the neutron number.
In previous studies \cite{DOTE,ENYObc} on Be isotopes, the excited 
states of $^{11}$Be were not 
studied in detail because the calculations are based on the variation 
before a total-angular-momentum projection.
Recently, the AMD framework has been developed to be an extended version
based on variational calculations after a spin-parity projection(VAP),
which has already been confirmed to be
powerful for studying the excited states of light nuclei.
The method has been applied to  a stable nucleus
 $^{12}$C \cite{ENYOe}, and to unstable nuclei:
$^{10}$Be, $^{12}$Be and $^{14}$Be\cite{ENYOf,ENYOg,ENYOh}.

In the present work, the structures of the ground and excited states 
of $^{11}$Be
were studied by performing VAP calculations using the AMD method. 
In the next section (Sec. \ref{sec:formulation}), 
we explain the formulation of AMD for a nuclear structure study of 
the ground and excited states. The adopted effective interactions 
are briefly explained  in Sec. \ref{sec:interaction}.  
In Sec.\ref{sec:results}, we present the theoretical results
concerning the energy levels, $\beta$ decays and
$E2$ transitions compared with the experimental data.
We predict a new rotational band with a well-developed clustering structure
which comes from $2\hbar\omega$ configurations.
In Sec.\ref{sec:discus}, the intrinsic structure 
and behavior of the valence neutrons are discussed.
We also discuss the breaking of $\alpha$-clusters which has an important
effect on the $\beta$-decay strength.
Finally, we summarize our work in Sec. \ref{sec:summary}

\section{Formulation}
 \label{sec:formulation}

In this section, the formulation of AMD 
for a nuclear structure study of the excited states
is explained briefly. For more detailed descriptions of the AMD framework
the reader is referred to Refs. \cite{ENYOg,ENYObc,ENYOe}

\subsection{Wave function}
An AMD wave function of a nucleus with mass number $A$
is a Slater determinant of Gaussian wave packets;
\begin{eqnarray}
&\Phi_{AMD}({\bf Z})=\frac{1}{\sqrt{A!}}
{\cal A}\{\varphi_1,\varphi_2,\cdots,\varphi_A\},\\
&\varphi_i=\phi_{{\bf X}_i}\chi_{\xi_i}\tau_i :\left\lbrace
\begin{array}{l}
\phi_{{\bf X}_i}({\bf r}_j) \propto
\exp\left 
[-\nu\biggl({\bf r}_j-\frac{{\bf X}_i}{\sqrt{\nu}}\biggr)^2\right],\\
\chi_{\xi_i}=
\left(\begin{array}{l}
{1\over 2}+\xi_{i}\\
{1\over 2}-\xi_{i}
\end{array}\right),
\end{array}\right. 
\end{eqnarray}
where the $i$th single-particle wave function $\varphi_i$
is a product of the spatial wave function $\phi_{{\rm X}_i}$,
 the intrinsic spin function $\chi_{\xi_i}$ and 
the isospin function $\tau_i$. 
The spatial part $\phi_{{\rm X}_i}$ is presented by 
a Gaussian wave packet whose center is defined by complex 
parameters $X_{1i}$, $X_{2i}$, $X_{3i}$.
$\chi_{\xi_i}$ is the intrinsic spin function parameterized by
$\xi_{i}$, while $\tau_i$ is the isospin
function which is fixed to be up(proton) or down(neutron)
 in the present calculations.
Thus an AMD wave function is parameterized by a set of complex parameters
${\bf Z}\equiv \{X_{ni},\xi_i\}\ (n=1,3\ \hbox{and }  i=1,A)$.
${\bf X}_{i}$ are the centers of Gaussians for spatial parts
and the parameters $\xi_{i}$'s determine the directions of  
intrinsic spins of the single particle wave functions.

If we consider a parity eigenstate projected from an AMD wave function,
the total wave function consists of two Slater determinants,
\begin{equation}
\Phi({\bf Z})=(1\pm P) \Phi_{AMD}({\bf Z}),
\end{equation}
where $P$ is a parity projection operator.
In the case of a total-angular-momentum projection($J$-projection),
the wave function of a system
is represented by the integral of the rotated states,
\begin{equation}
\Phi({\bf Z})=P^J_{MK'}\Phi_{AMD}({\bf Z}) = 
\int d\Omega D^{J*}_{MK'}(\Omega)R(\Omega)\Phi_{AMD}({\bf Z}),
\end{equation}
where the function $D^J_{MK}$ is
the well-known Wigner's D function and $R(\Omega)$
stands for the rotation operator with Euler angle $\Omega$.

In principal the total wave function can be a
superposition of independent AMD wave functions. 
For example, a system is written by a superposition 
of spin-parity projected AMD wave functions $P^{J\pm}_{MK'}\Phi_{AMD}$
as follows, 
\begin{equation} \label{eqn:superpose}
\Phi=cP^{J\pm}_{MK'}\Phi_{AMD}({\bf Z})
+c'P^{J\pm}_{MK'}\Phi_{AMD}({\bf Z}')+\cdots.
\end{equation}

 Expectation values of a given tensor operator $T^k_q$(rank $k$) for the 
total-angular-momentum projected states 
$P^{J_1\pm}_{M_1K_1}\Phi_{AMD}({\bf Z})$
and $P^{J_2\pm}_{M_2K_2}\Phi_{AMD}({\bf Z'})$ are calculated as follows,
\begin{eqnarray}
&&\langle P^{J_1}_{M_1K_1}\Phi_1|T^k_q|P^{J_2}_{M_2K_2}\Phi_2\rangle \\
&&=\frac{2J_2+1}{8\pi^2} (J_2M_2kq|J_1M_1)
\sum_{K\nu}(J_2Kk\nu |J_1K_1) \int d\Omega D^{J_2*}_{KK_2}(\Omega)
\langle\Phi_1|T^k_\nu R(\Omega)|\Phi_2\rangle.
\end{eqnarray}  
The three-dimensional integral can be evaluated 
numerically by taking a finite number of mesh points of the Euler angle
$\Omega$=($\alpha,\beta,\gamma$).

\subsection{Energy variation}
We make variational calculations  for a trial wave function
to find the state which 
minimizes the energy of the system;
\begin{equation}
\frac{\langle\Phi|H|\Phi\rangle}{\langle\Phi|\Phi\rangle}.
\end{equation}
In the AMD framework, the energy variation is performed 
by a method of frictional cooling, one of the imaginary time methods.
Regarding the frictional cooling method, the reader is referred to 
papers \cite{ENYOa,ENYObc}.
The time development of the parameters ${\bf Z}$ of 
a wave function $\Phi({\bf Z})$ 
is simulated by the frictional cooling equations,
\begin{equation}
\frac{dX_{n k}}{dt}=
(\lambda+i\mu)\frac{1}{i \hbar} \frac{\partial}{\partial X^*_{n k} }
\frac{\langle \Phi({\bf Z})|H|\Phi({\bf Z})\rangle}{\langle \Phi({\bf Z})
|\Phi({\bf Z})\rangle},
\quad (n=1,3\quad k=1,A)
\end{equation}
\begin{equation}
\frac{d\xi_{k}}{dt}=(\lambda+i\mu)\frac{1}{i\hbar}
\frac{\partial}{\partial\xi^*_{k}}
\frac{\langle \Phi({\bf Z})|H|\Phi({\bf Z})\rangle}{\langle \Phi({\bf Z})
|\Phi({\bf Z})\rangle},
\quad (k=1,A)
\end{equation}
with arbitrary real numbers $\lambda$ and $\mu < 0$. It is easily proved that 
 the energy of the system decreases with each time step due to the 
frictional term:$\mu$. After sufficient cooling iterations,
 the parameters for 
the minimum-energy state are obtained.

\subsection{Wave function for $J^\pm$ states}
In order to obtain the wave function for a $J^\pm$ state,
we make the energy variation after a spin-parity projection(VAP)
for an AMD wave function
by using the frictional cooling method explained above.
That is to say we perform the energy variation for the trial function
$\Phi=P^{J\pm}_{MK'}\Phi_{AMD}({\bf Z})$,
the spin-parity eigenstate projected
from an AMD wave function. 
First we make variational calculations after only the parity projection
but before the spin projection(VBP)
to prepare an initial trial wave function
 $\Phi_{AMD}({\bf Z}_{init})$.
After obtaining an initial wave function in VBP calculations
we  evaluate the expectation 
values of Hamiltonian 
for the spin-parity-projected states by choosing 
the body-fixed $3$-axis for the $\Omega$ rotation
to be the approximate principal $z$-axis on the intrinsic deformation.
Then we find an appropriate $K'$ quantum that gives the minimum 
diagonal energy of the spin-parity eigenstate
\begin{equation}
\langle P^{J\pm}_{MK'}\Phi_{AMD}({\bf Z}_{init})|H|P^{J\pm}_{MK'}
\Phi_{AMD}({\bf Z}_{init}) \rangle \over 
\langle P^{J\pm}_{MK'}({\bf Z}_{init})|P^{J\pm}_{MK'}({\bf Z}_{init})
 \rangle,
\end{equation}
where $K'=\langle J_3\rangle$.
For each spin parity $J^\pm$, we start VAP calculations 
for the normalized energy expectation value
$\langle P^{J\pm}_{MK'}\Phi_{AMD}({\bf Z})|H|P^{J\pm}_{MK'}
\Phi_{AMD}({\bf Z}) \rangle/ 
\langle P^{J\pm}_{MK'}({\bf Z})|P^{J\pm}_{MK'}({\bf Z})
 \rangle$ with the adopted $K'$
quantum from the initial state.
In general, the direction of the approximately principal $z$-axis 
is automatically determined in the energy variation because
the shape of the intrinsic system
can vary freely.
The approximately principal axis
can deviate from the $3$-axis in the VAP procedure with 
a given $K'=\langle J_3\rangle$.
That is to say that the optimum state $P^{J\pm}_{MK'}\Phi_{AMD}$ obtained after
the variation may contain so-called $K$-mixing ($K=\langle J_z\rangle$) 
components.
However, the deviation of $z$-axis from the $3$-axis 
is found to be small in many cases.
It means that the obtained states do not contain $K$-mixing components
 so much, and $K'$ well corresponds to the $K$
quantum.

Concerning the states in the lowest band with each parity, we can obtain
appropriate initial wave functions by simple VBP calculations as mentioned
above.  For the highly excited states,
in order to obtain initial wave functions and appropriate $K'$ quanta
for VAP calculations
we make VBP calculations with a constraint on AMD wave function.
The details of the AMD calculation with constraints have been described 
in Refs. \cite{ENYOa}. In the present calculations, we adopt a constraint 
as the expectation value of the total-oscillator quanta to equal with 
a given number. After choosing a corresponding $K'$
quantum, we perform VAP calculations from the initial wave function.
When we obtain other 
local minimum states than the obtained states,
we consider them as the states in higher rotational bands.

\subsection{Diagonalization}  

After the VAP calculations for $J^\pm_n$ states, 
the optimum intrinsic states
$\Phi^1_{AMD}, 
\Phi^2_{AMD},\cdots$, and  
$\Phi^m_{AMD}$ are obtained. Here $m$ indicates the number
of the calculated levels.
we consider that the obtained wave functions approximately represent
the intrinsic wave functions of the $J^\pm_n$ states.
We determine the final wave functions by superposing the obtained
AMD wave functions. That is to say that we determine the coefficients 
$c,c',\cdots$
in Eq. \ref{eqn:superpose} for each $J^\pm_n$ state
by diagonalizing the Hamiltonian matrix 
$\langle P^{J\pm}_{MK'} \Phi^i_{AMD}
|H|P^{J\pm}_{MK''} \Phi^j_{AMD}\rangle$
and the norm matrix
$\langle P^{J\pm}_{MK'} \Phi^i_{AMD}
|P^{J\pm}_{MK''} \Phi^j_{AMD}\rangle$
simultaneously with regard to ($K', K''$) and ($i,j$).
In comparison with the experimental data such as the energy levels and
the strength $E2$ transitions, the theoretical values are calculated with the 
final states after diagonalization.

\section{Interactions} 
\label{sec:interaction}

The adopted interaction for the central force is the case 3 of 
MV1 force \cite{TOHSAKI},
which contains a zero-range three-body term:$V^{(3)}$ 
in addition to the two-body interaction:$V^{(2)}$, 
\begin{eqnarray}
& V_{DD}=V^{(2)}+V^{(3)}\\
& V^{(2)}=\sum_{i<j}
(1-m+b P_\sigma-h P_\tau -m P_\sigma P_\tau )
\left\lbrace 
V_A \exp\left[-\left(\frac{|{\bf r}_i-{\bf r}_j|}{r_A}\right)^2\right]+ 
V_R \exp\left[-\left(\frac{|{\bf r}_i-{\bf r}_j|}{r_R}\right)^2\right]
\right\rbrace,\\
& V^{(3)} =\sum_{i<j<k} v^{(3)}\delta({\bf r}_i-{\bf r}_j)
\delta({\bf r}_i-{\bf r}_k),
\end{eqnarray}
where $P_\sigma$ and $P_\tau$ stand for the spin and isospin exchange 
operators, respectively. 
As for the two-body spin-orbit force:$V_{LS}$, we use the G3RS force 
\cite{LS} as follows,
\begin{eqnarray}
& V_{LS}= \sum_{i<j}\left\{ u_I \exp\left(-\kappa_I r^2\right) +
u_{II} \exp\left(-\kappa_{II} r^2\right)\right\} 
\frac{(1+P_\sigma)}{2}
\frac{(1+P_\tau)}{2}
{\bf l}\cdot ({\bf s}_i+{\bf s}_j),\\
\end{eqnarray}
The Coulomb interaction:$V_{{\rm C}}$ 
is approximated by a sum of seven Gaussians.
The total interaction V is a sum of these interactions:
$V=V_{DD}+V_{LS}+V_C$. 

\section{Results}\label{sec:results}

The structures of the excited states of $^{11}$Be are studied 
based on the VAP calculations in the framework of AMD.
In this section we present the theoretical results concerning
the energy levels, the $E2$ transitions, and the $\beta$ transitions,
which should be directly compared with the experimental data.
More detailed discussions of the intrinsic structures are
given in the next section.

We adopt two sets of the interaction parameters.
One parameter set (1) is $m=0.65$, $b=h=0$ for Majorana,
Bartlett and Heisenberg terms in the central force and 
$u_I=-u_{II}=3700$ MeV for the strength 
of the spin-orbit forces, which have been used in the
previous study on the excited states of  $^{10}$Be.
We also try another set (2) with weaker spin-orbit forces as
$u_I=-u_{II}=2500$ MeV. Other parameters in the case (2) are same as those in
the case (1). The width parameter $\nu$ is chosen to be 0.18 fm$^{-2}$ 
which gives the minimum energy of $^{11}$Be in a VBP calculation.

In VBP calculations we know that the lowest positive parity band 
$K^\pi=1/2^+$consists of $1/2^+$,
$3/2^+$,$5/2^+$,$7/2^+$ and $9/2^+$ states, and the lowest negative parity
band $K^\pi=1/2^-$ consists of $1/2^-$, $3/2^-$, and $5/2^-$ states.
Therefore the $J^\pm_1$ states in the lowest bands are obtained by VAP 
calculations for $P^{J\pm}_{MK'}\Phi_{AMD}$ with
the corresponding $(J^\pm,K')$ values as 
$(1/2^+,1/2)$, $(3/2^+,1/2)$, $(5/2^+,1/2)$, $(7/2^+,1/2)$,
$(9/2^+,1/2)$, $(1/2^-,1/2)$, $(3/2^-,1/2)$, $(5/2^-,1/2)$.
We calculate the higher excited states in the second negative parity band
by VAP calculations with 
$(J^\pm,K')$=
$(7/2^-,3/2)$, $(9/2^-,3/2)$, $(11/2^-,3/2)$, $(13/2^-,3/2)$, $\cdots$.
These states are considered to belong to a band with $K^\pi=3/2^-$ . 
After obtaining the intrinsic states in this second negative parity band
for $J\ge 7/2$,
the excited  $J^\pm=3/2^-_2$ and $5/2^-_2$ states in $K^\pi=3/2^-$ 
are found as local minimums with VAP calculations by starting from the 
intrinsic states obtained for the higher spin states.
In order to find the other excited $3/2^-$ state, 
we make VAP calculations for the spin-parity projected AMD wave function 
with the fixed intrinsic spin directions as 3 spin-down protons, one spin-up
proton, 4 spin-down and 4 spin-up neutrons.
The obtained wave function for the $3/2^-$ state 
is dominated by a component of total-intrinsic spin $S_p=1$ for protons.
We superpose the wave functions to calculate the final wave functions by 
diagonalizing Hamiltonian and norm matrixes simultaneously.

We should notice that all possible excited states are 
not exhausted in the present calculations. In this work we perform 
VAP calculations basically for the rotational states which can be known 
from VBP calculations with or without constraint. For the higher excited 
states we must extend the VAP calculations
with orthogonal condition to the lower states by
superpositions as the previous studies \cite{ENYOg,ENYOe}.
Above the states shown in the present results, there should exist 
 other excited states which may be obtained with the 
extended VAP calculations.

\subsection{Energies}

The theoretical binding energies of $^{11}$Be are 58.2 MeV in the case (1)
and 54.4 MeV in the case (2),
 both of which underestimate the experimental value 65.48 MeV. 
The binding energy can be reproduced by choosing a interaction 
parameter such as $m=0.60$ of
the Majorana term. However unfortunately it is difficult to
reproduce all the features of nuclear structures such as 
binding energies, energy levels, radii, deformations and so on 
with one set of interaction parameters.
Since we study the excited states taking care about the excitation energies 
and the intrinsic structures, we adopt the case(1) interaction which 
reproduces well the features of the excited states of $^{10}$Be\cite{ENYOg}
except for the binding energy. With this interaction 
the spin-parity $1/2^+$ of the ground state of $^{11}$Be can be described
in the present calculations.
We also use another interaction case (2) with the weaker spin-orbit 
forces to be compared.
The improvement of effective interactions is one of the important 
problems in nuclear studies.

Here we comment on the stability of AMD wave functions
above threshold energies. Since the single particle wave functions are
written by Gaussians in AMD calculations, 
the relative motion between particles in a system
is restricted by a Gaussian or a linear combination of Gaussians.
Because of the limitation of the model space,
continuum states nor out going waves can not be represented in the 
present model. Even if the energy of a nucleus is above the 
threshold energies of particle decays, the particles can not 
necessarily go away in the present framework. In this sense, the
system is treated in the bound state approximation.
The widths for the particle decays should be
discussed carefully by other frameworks such as the method with
reduced width amplitude or the complex scaling method
beyond the present AMD framework.

\begin{figure}
\noindent
\epsfxsize=0.4\textwidth
\centerline{\epsffile{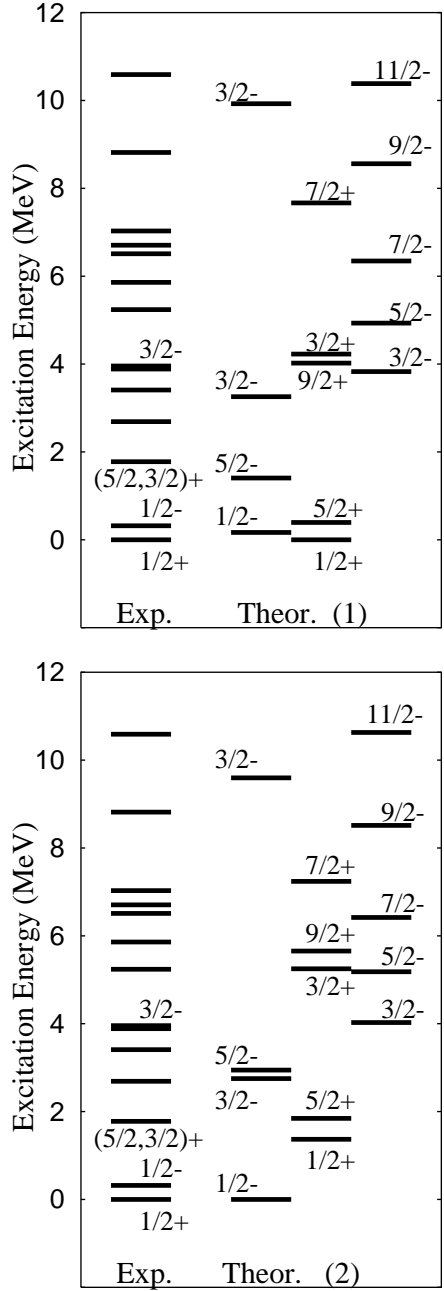}}
\caption{\label{fig:be11spe}
The excitation energies of the excited states of $^{11}$Be.
Theoretical results with the case(1) and case(2) interactions are compared with
the experimental data quoted from 
`Table of Isotopes'.}
\end{figure}

\begin{figure}
\noindent
\epsfxsize=0.45\textwidth
\centerline{\epsffile{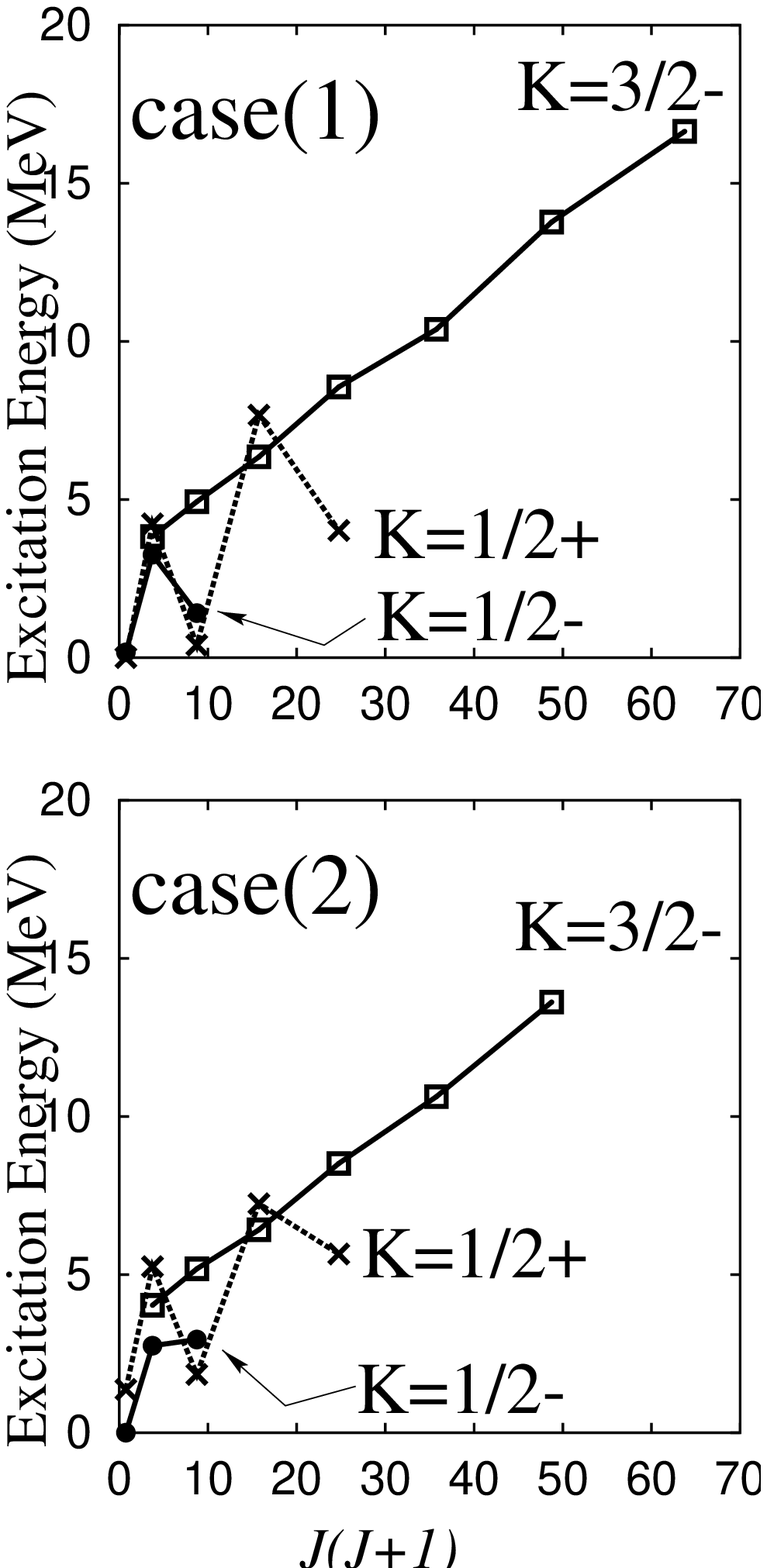}}
\caption{\label{fig:be11rot}
The excitation energies of the rotational bands of $^{11}$Be
as a function of $J(J+1)$. The lines indicate the theoretically 
obtained rotational bands $K^\pi=1/2^-$,  $K^\pi=1/2^+$,  
$K^\pi=3/2^-$.}
\end{figure}

The energy levels of $^{11}$Be are shown in Fig.\ref{fig:be11spe}.
There are many low-lying levels in the experimental data.
The abnormal spin parity $1/2^+$ of the ground state has been known,
although the normal spin parity of $^{11}$Be is $1/2^-$ in 
the simple shell-model picture.
The calculations with the case(1)
interaction reproduce the parity inversion between $1/2^+$ and $1/2^-$
states.
In the theoretical results, the rotational bands $K^\pi=1/2^+$ and
 $K^\pi=1/2^-$ start from the band head 
$1/2^+$ and $1/2^-$ states, respectively.
Also in the results with the case(2) interaction, the $1/2^+$ state is 
lowest in the $K^\pi=1/2^+$ band, however, it is slightly higher
than the negative parity $1/2^-$ state by 1.4 MeV.
As a result, 
the abnormal parity of the ground state can not be reproduced by the case (2) 
interaction which the spin-orbit forces are weaker than those in the case (1).
We should not conclude the calculations with the case(1) interaction 
are better than those in the case (2), because the neutron halo effect 
on the energy gain of $1/2^+$ state is not 
taken into account in the present calculations. The details of the 
reproduction of parity inversion 
with the case(1) interaction are described in the later section.

In both results in the case (1) and the case (2), there exist some rotational 
bands in the low energy region. By classifying the calculated excited states
we can obtain three rotational bands $K=1/2^+$, $K=1/2^-$ and $K=3/2^-$
which are dominated by $1\hbar\omega$, $0\hbar\omega$ and $2\hbar\omega$
neutron configurations, respectively.  
In Fig.\ref{fig:be11rot}, we show the excitation energies of the rotational
bands as a function of total spin $J(J+1)$.
We find a new eccentric band 
$K=3/2^-$ which starts from the second $3/2^-$ state at about 4 MeV.
The excited states  in the $K=3/2^-$ band are dominated by $2\hbar\omega$ 
excited configurations with 2 particles and 3 holes
in neutron shells.
The $K=3/2^-$ band is constructed by a well-deformed 
intrinsic state with a developed clustering structure. The rotational band 
indicates large moment of inertia and it reaches the high spin states 
about 20 MeV. 
The highest spin is $15/2^-$ in the case (1) and
$13/2^-$ in the case (2). In VAP calculations with further high spins,
we can not obtain any stable states 
since an $\alpha$ particle escapes far away in the variational calculations.
It means that the energy of the relative motion between clusters
is beyond the classical barrier due to the coulomb and centrifugal forces.
Since the barrier hight is very sensitive to the binding energy,
we examine the highest spin of the $K=3/2^-$ band 
with another set of the interaction parameters taking care 
of the binding energy
of $^{11}$Be.
If we change the Majorana exchange term of case (1) to 
 $m=0.60$, we can make the system bound as deeply as the experimental 
binding energy. Then a $17/2^-$ state is obtained as the highest spin state.
W. Von Oertzen et al. suggested some candidates 
for the states belonging to this new negative parity band in 
the experimental data
observed in the 2 neutron transfer reactions,
$^9$Be($^{13}$C, $^{11}$C)$^{11}$Be \cite{OERTZENa}.
Although they suggested a possibility of $19/2^-$ state, 
the present results are negative to such a high spin 
state as $19/2^-$ in the rotational $K=3/2^-$ band,
because $17/2^-$ is the highest spin made from the $2p$-$3h$ configurations.
It should be noticed that the present work is the first microscopic 
calculation which predicts the $K^\pi=3/2^-$
band with the well-developed clustering structure.
Since $^{11}$Be is a loosely bound system, further studies 
taking account of widths of the 
excited states are necessary to determine the band terminal.

We find a excited $3/2^-$ state at about 10 MeV
 (Fig \ref{fig:be11spe}).
In this state, the proton structure is quite 
strange comparing with the other excited states of 
$^{11}$Be.
In most of the states of $^{11}$Be, we find $2\alpha$ cores 
as well as other Be isotopes: $^8$Be, $^9$Be and $^{10}$Be.
However, in the $3/2^-$ state at 10 MeV,
 only one $\alpha$-cluster is formed. The other 2 protons do not form
a $\alpha$-cluster but couple to be totally $S=1$ with 
aligned intrinsic spins.
It is analogue with the structure of the first $1^+$ state with 
the unnatural spin parity in $^{12}$C. We consider that this $3/2^-$
state must be the newly measured state at 8.04 MeV \cite{AOI} 
to which the $\beta$-decay transition from $^{11}$Li is strong.
Although its excitation energy is overestimated in the present calculations,
it is easily improved by changing parameters $b$ and $h$ of 
the Bartlett and Heisenberg terms. For example the parameters 
$b=-0.2$, $h=0.4$, $m=0.41$ gives 2 MeV lower excitation 
energy of the $3/2^-$ state than 
the one with $b=0.0$, $h=0.0$, $m=0.65$. Here we fit the Majorana 
parameter $m$ as to give the same $\alpha$-$\alpha$ interaction as the 
case (1). The change of parameters gives no significant 
effect on the excitation energies of the other states with 2-$\alpha$ clusters.
The results indicate that the Bartlett and 
Heisenberg terms should be taken into consideration
 in the detailed study of the energy levels though 
they have been often omitted in the traditional works on stable nuclei.

\subsection{Transition strength}

Data concerning the $\beta$-decay strength are very useful to investigate
structures of excited states.
There are many experimental data concerning the $\beta^-$ and $\beta^+$ decays
into the excited states of $^{11}$Be. 
The strength of the $\beta^-$ decays from $^{11}$Li has 
been measured recently
\cite{AOI}. For the $\beta^+$ decays, the strength of Gamov-Teller(GT)
 transitions has been deduced from the charge exchange reactions 
$^{11}$B$(t,^3$He)$^{11}$Be \cite{FUJIWARA}.
In the GT transitions from 
$^{11}$Li$(3/2^-)$ and $^{11}$B$(3/2^-)$, 
the allowed daughter states are $1/2^-$, $3/2^-$ or $5/2^-$ states.  
In Table \ref{tab:be11beta} the experimental log$(ft)$ values are presented
comparing with the theoretical results.
In order to calculate the log($ft$) values of GT transistions into the 
excited states of $^{11}$Be we prepare the parent states   
$^{11}$Li$(3/2^-)$ and $^{11}$B$(3/2^-)$ by VAP calculations in the  
same framework.

The experimental log($ft$) values are reproduced well by the theoretical
calculations.
It is easily understood that the decays into the excited states in
$K=3/2^-$ band are weak because these states 
have the ell-developed clustering structures with dominated 
$2\hbar\omega$ components
which make the overlap of a GT operator with the parent
 state of $^{11}$B to be small.
On the other hands, 
we consider that three levels at 0.32 MeV, 2.69 MeV and  
3.96 MeV measured in 
the $\beta^+$ transitions correspond to the states $1/2^-_1$, $3/2^-_1$ and 
$5/2^-_1$ in the $K=1/2^-$ band because
the experimental data indicate the significant $\beta^+$ strength 
as log$(ft)$ $<$ 5.0, which consistent with the theoretical results.

What is important in the $\beta^-$ decays from $^{11}$Li
is that the strength is very sensitive to the breaking of 
2 $\alpha$-cluster cores in $^{11}$Be.
If the daughter state of $^{11}$Be poses 2 ideal $\alpha$-cluster 
cores with $(0s)^4$ configurations, the GT transitions from $^{11}$Li 
are forbidden completely because of Pauli principle. In another word, the
strength of GT transitions from $^{11}$Li indicates the degree of 
the $\alpha$-cluster breaking in the daughter states of 
$^{11}$Be. 
From this point of view, one of the reasons for the weak
$\beta^-$ transitions as log$(ft)$=5.67 to the lowest $1/2^-$ state 
at 0.32 MeV is the 
2$\alpha$-cluster structure in $^{11}$Be($1/2^-$).
Another reason for the weak $\beta^-$ transitions has been suggested to be
the effect of the halo structure in $^{11}$Li by T. Suzuki et al. 
\cite{TSUZUKIa}.
The ground state of $^{11}$Li is known to have the neutron halo structure 
which originates from $s$-orbits. The mixing of the $s$-orbits 
causes weak $\beta^-$ transitions to the normal states of $^{11}$Be.
 Since the halo structure of $^{11}$Li 
can not described in the present AMD wave function, the possible halo effects
on the $\beta^-$ decays are not included in the present results.
If we suppose the mixing ratio of $s$-orbits in the ground $^{11}$Li state as
to be 50 \%, the theoretical log$(ft)$ values concerning the
$\beta^-$ decays into $1/2^-_1$, $3/2^-_1$ and $5/2^-_1$ is expected to
be increased by log(2)$\sim 0.3$ due to the mixing.
Even if we add 0.3 of the halo effect by hands to the present
log($ft$) values for $3/2^-_1$ and $5/2^-_1$ states, the log($ft$) values
in case(1) are still smaller than 5.0, because of the sufficient cluster 
breaking.
They are consistent with the experimentally 
measured rather small log$ft$ values for the decays to $^{11}$Be(2.7 MeV)
and $^{11}$Be(3.9 MeV). In another word,
the experimental data indicates the significant breaking of 
2 $\alpha$-cluster cores in these states(2.7MeV and 3.9MeV) which we 
consider as the excited states 
$3/2^-_1$ and $5/2^-_1$.

In the recent experiments of the measurements of $\beta^-$ decays,
a new excited state at 8.04 MeV  with strong $\beta^-$ transitions
 has been discovered. The calculated log$(ft)$ 
values for the $3/2^-$ state of $^{11}$Be at about 10 MeV well corresponds 
to this newly observed state at 8.04 MeV.
The transition is strong because the 2$\alpha$-cluster structure disappears
completely in this state.

As shown in Table \ref{tab:be11beta}, the experimental data concerning the
strength of $\beta^-$ decays have been systematically reproduced in 
the present calculations. 
Considering the increases of log$ft$ 
due to the halo effect of $^{11}$Li,
the theoretical values of 
log$ft$ for $\beta^-$ calculated with the case (1) interaction 
well agree to the experimental data. 
The reason for the good reproduction 
is because the significant breaking 
of 2-$\alpha$ clusters in $^{11}$Be is described in the present calculations.
It is the same reason as the previous studies of $^{12}$C in which 
the experimental data concerning the strength of $\beta^-$ and $\beta^+$
 decays have been well reproduced. 
The quantitative discussion of the 2$\alpha$-core breaking
will be given in the next section. 

In Table \ref{tab:be11be2}, we show the theoretical $B(E2)$ values.
It has been proved that AMD calculations reproduce well $B(E2)$ values
of light neutron-rich nuclei.
As shown in the previous studies \cite{ENYOf,ENYObc,ENYOe},
the experimental $Q$-moments and $B(E2)$ values of various nuclei have been 
reproduced well by using the bare charges in the AMD framework
because of the advantage of the AMD method which 
can directly describe proton deformations.
Also in the present results, the features of proton-matter deformations 
in the intrinsic states are reflected in the theoretical 
$B(E2)$ values.
In the lowest negative parity band $K^\pi=1/2^-$,
the intrinsic system deforms because of 2 
$\alpha$-cluster cores although the development of clustering is smallest
comparing with the other bands.  
$B(E2)$ values are larger in the ground $K^\pi=1/2^+$ band which has 
the developed clustering structure with a large deformation.
The $B(E2)$ values are enhanced between the states in the
second negative parity band $K^\pi=3/2^-$ with an extremely large deformation
due to the clustering development.

\begin{table}
\caption{\label{tab:be11beta} Log(ft) values of the $\beta$ transitions.
The theoretical values are obtained from the Gamow-Teller transition strength.
The experimental data of $\beta$ decays from $^{11}$Li are taken from
Ref.\protect\cite{AOI}. Log$ft$ values concerning the
 Gamow-Teller transition from $^{11}$B
are deduced from the charge exchange reactions \protect\cite{FUJIWARA}. 
}

\begin{center}
\begin{tabular}{ccccc}
 transitions &   log$ft$ & \\
    &   theory (1) & theory (2) \\
\hline
$^{11}$Li($3/2^-$) $\rightarrow ^{11}$Be($1/2^-_1$) & 5.0 & 5.5  \\
$^{11}$Li($3/2^-$) $\rightarrow ^{11}$Be($3/2^-_1$) & 4.4 & 5.0  \\
$^{11}$Li($3/2^-$) $\rightarrow ^{11}$Be($5/2^-_1$) & 4.5 & 5.0  \\
$^{11}$Li($3/2^-$) $\rightarrow ^{11}$Be($3/2^-_2$) & 4.9 & 6.2  \\
$^{11}$Li($3/2^-$) $\rightarrow ^{11}$Be($3/2^-_3$) & 3.9 & 4.3  \\
\hline
$^{11}$B($3/2^-$) $\rightarrow ^{11}$Be($1/2^-_1$) & 3.9 & 4.2  \\
$^{11}$B($3/2^-$) $\rightarrow ^{11}$Be($3/2^-_1$) & 3.8 & 4.3  \\
$^{11}$B($3/2^-$) $\rightarrow ^{11}$Be($5/2^-_1$) & 4.2 & 5.0  \\
$^{11}$B($3/2^-$) $\rightarrow ^{11}$Be($3/2^-_2$) & 4.5 & 5.5  \\
$^{11}$B($3/2^-$) $\rightarrow ^{11}$Be($3/2^-_3$) & 4.3 & 5.4   \\
\hline
 transitions  &  exp. &   \\
$^{11}$Li($3/2^-$) $\rightarrow ^{11}$Be($1/2^-_1$, 0.32 MeV) & 5.67(4) &  \\
$^{11}$Li($3/2^-$) $\rightarrow ^{11}$Be(2.69 MeV) & 4.87(8) &  \\
$^{11}$Li($3/2^-$) $\rightarrow ^{11}$Be(3.96 MeV) & 4.81(8) &  \\
$^{11}$Li($3/2^-$) $\rightarrow ^{11}$Be(5.24 MeV) & 5.05(8) &  \\
$^{11}$Li($3/2^-$) $\rightarrow ^{11}$Be(8.04 MeV) & 4.43(8) &  \\
\hline
 transitions  &  exp. &   \\
$^{11}$B($3/2^-$) $\rightarrow ^{11}$Be($1/2^-_1$, 0.32 MeV) & 4.3(1) &  \\
$^{11}$B($3/2^-$) $\rightarrow ^{11}$Be(2.69 MeV) & 4.4(1) &  \\
$^{11}$B($3/2^-$) $\rightarrow ^{11}$Be(3.96 MeV) & 4.8(2) &  \\
\end{tabular}
\end{center}
\end{table}

\begin{table}
\caption{\label{tab:be11be2} $E2$ strength of $^{11}$Be. The 
theoretical results of AMD with the interactions case(1) are listed.
}

\begin{center}
\begin{tabular}{ccccc}
 transitions &  present results (e$^2$ fm$^4$)\\
\hline
$^{11}$Be;$3/2^-_1\rightarrow 1/2^-_1$ & 9 \\
$^{11}$Be;$5/2^-_1\rightarrow 1/2^-_1$ & 8 \\
$^{11}$Be;$5/2^-_1\rightarrow 3/2^-_1$ & 2 \\
\hline
$^{11}$Be;$5/2^-_2\rightarrow 3/2^-_2$ & 37 \\
$^{11}$Be;$7/2^-_1\rightarrow 3/2^-_2$ & 14 \\
$^{11}$Be;$7/2^-_1\rightarrow 5/2^-_2$ & 25 \\
\hline
$^{11}$Be;$5/2^+_1\rightarrow 1/2^+_1$ & 14 \\
$^{11}$Be;$5/2^+_1\rightarrow 3/2^+_1$ & 7 \\
$^{11}$Be;$3/2^+_1\rightarrow 1/2^+_1$ & 13 \\
\hline
$^{11}$Be;$5/2^-_2\rightarrow 3/2^-_1$ & 7 \\
$^{11}$Be;$3/2^-_2\rightarrow 3/2^-_1$ & 8 \\
$^{11}$Be;$3/2^-_2\rightarrow 1/2^-_1$ & 2 \\
\end{tabular}
\end{center}
\end{table}

The $E1$ transition strength from $1/2^-_1$ to $ 1/2^+_1$  in $^{11}$Be
is known to be very large comparing with those in other light nuclei. 
The theoretical value of $B(E1;1/2^-_1\rightarrow 1/2^+_1)$=0.02 e$\cdot$fm
is much smaller than the experimentally observed strength 
$0.116 \pm 0.011$ e$\cdot$fm. The halo structure may give an important 
effect on the $E1$ strength.

\section{Discussions}\label{sec:discus}

We investigate the structures of $^{11}$Be focusing on 
the clustering aspects. 
It is found that two $\alpha$-cluster cores
are formed in most of the states of $^{11}$Be as well as 
in other Be isotopes:$^8$Be, 
$^9$Be, $^{10}$Be. 
In the present calculations, we found three rotational bands with 
2-$\alpha$ cores in $^{11}$Be.
In this section,
we argue the development  of clustering 
and investigate the roles of valence neutrons in the clustering states. 
We analyze the breaking of cluster cores and also describe the 
feature of a non-clustering state. The problem of parity inversion 
between positive and negative parity states is also discussed.

\subsection{Clustering structure}

We investigate the intrinsic structures of the excited states focusing on the 
clustering aspects. Although
the states $P^{J\pm}_{MK}\Phi^i_{AMD}$
are superposed for all the obtained wave functions $\Phi^i_{AMD}$
so as to diagonalize a Hamiltonian matrix,
the $J^\pm$ state after the diagonalization is found to be dominated 
by the AMD wave function $P^{J\pm}_{MK}\Phi^j_{AMD}$ which is 
obtained in a VAP calculation for the given spin and parity $J^\pm$.
Therefore we consider an obtained AMD wave function $\Phi_{AMD}$ with
a VAP calculation as the intrinsic wave function of the corresponding 
$J^\pm$ state.

\begin{figure}
\noindent
\epsfxsize=0.45\textwidth
\centerline{\epsffile{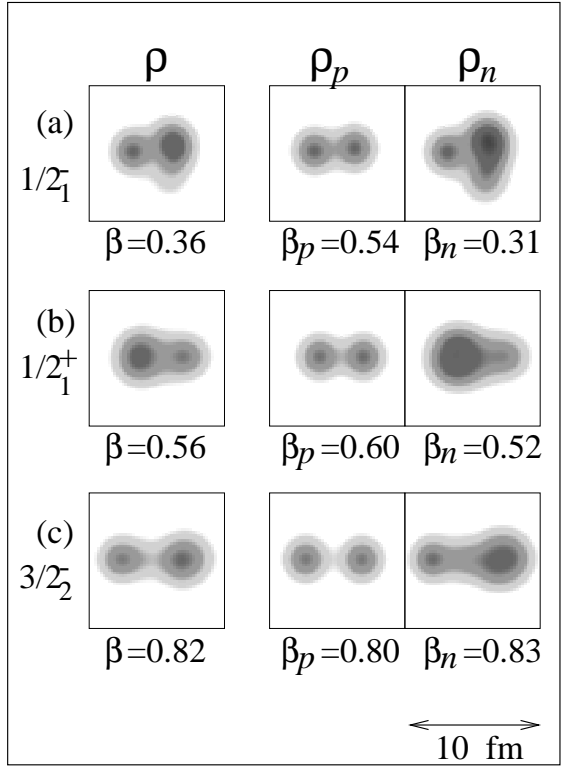}}
\caption{\label{fig:be11dens}
The density distributions of the intrinsic structures of 
$1/2^-_1$, $1/2^+_1$, $3/2^-_2$ states
 calculated with the case(1) interaction. 
The intrinsic system is projected on to a plain which contains
the longitudinal axis of the intrinsic states.
The density is integrated along a transverse axis perpendicular to the
plain.
The density distributions for matter, proton and neutron are 
presented in the left, middle 
and right, respectively.}
\end{figure}

In the excited states, three rotational bands:
$K=1/2^+$, $K=1/2^-$ and $K=3/2^-$ are recognized,
because the intrinsic structures of the states in each band are 
similar.
The density distributions of the 
intrinsic states of the band head states:
$1/2^-_1$, $1/2^+_1$ $3/2^-_2$ are shown in Fig.\ref{fig:be11dens}.
The neutron structure of an intrinsic state is very different from the
ones of the other rotational bands.
The density distributions of neutrons are presented in the right column in
Fig.\ref{fig:be11dens}. As for the neutron deformation, the $1/2^-$ state 
has an oblate deformation.
The oblate shape is natural for the normal parity 
state in a system with the neutron number $N=7$.
In the positive parity $1/2^+$ state, the neutron density deforms prolately. 
The prolate deformation of neutron density is extremely enhanced
in the $3/2^-_2$ state.
The detailed discussions of neutron structures is given later
by analyzing the single-particle behavior of neutrons. 

In the proton density shown in the middle column in Fig.\ref{fig:be11dens},
we can see dumbbell shapes due to 2 pairs of protons. 
Roughly speaking, it indicates that 2-$\alpha$ cores are formed 
in all the intrinsic states of these three rotational bands. 
The spatial development of clusters is small in the lowest negative parity 
$1/2^-_1$ state which has a main component of a 
normal $0\hbar\omega$ configuration.
In the ground $1/2^+$ state with the abnormal parity,
the $2\alpha$ clustering structure develops. It is interesting 
that the most remarkable
clustering structure is seen in the $3/2^-_2$ state which belongs to the 
$K=3/2^-$ band.
In this state, the relative distance between
2-$\alpha$ cores is largest among these 3 rotational bands.
It seems that the $2\alpha$-clustering structure develops following the
prolate deformation of neutron density.
According to the development of $2\alpha$ clustering,
the proton deformation is larger in the $K=1/2^+$ band than 
in the $K=1/2^-$ band, and largest in the $K=3/2^-$ band. The difference 
of the proton structures are reflected in the theoretical $B(E2)$ values 
as mentioned in the previous section. Namely, the theoretical values 
concerning the strength $B(E2)$ are large in the transitions between
the states in the rotational bands with the large proton deformations.
As mentioned above, the $K=3/2^-$ band is the largely deformed state 
with the well development clustering structure. 
However, the deformation decreases and the clustering weakens at  
the band terminal state with the increase of spin 
near the highest spin $J=15/2^-$ in case(1).

\subsection{Behavior of valence neutrons}

As mentioned in the previous subsection,
the deformation of neutron density 
changes drastically between the rotational bands.
We inspect the single-particle wave functions of the intrinsic states
to study the behavior of valence neutrons.
The single-particle wave functions of an intrinsic state
are obtained by transforming from the original Gaussian functions to 
the orthonormal single-particle wave functions under the diagonal condition 
of the single-particle Hamiltonian matrix. The definition of single-particle
wave functions and energies of an AMD wave function is described
in Refs. \cite{DOTE,ENYOg}.

\begin{figure}
\noindent
\epsfxsize=0.49\textwidth
\centerline{\epsffile{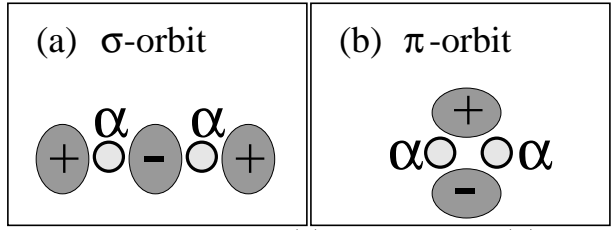}}
\caption{\label{fig:sigmapi}
Sketches for the molecular orbits, (a) $\sigma$-orbits and (b)$\pi$-orbits 
surrounding $2\alpha$s. The molecular orbits are understood by linear
combinations of $p$-shell orbits around the $\alpha$ clusters.
}
\end{figure}

In Be isotopes, it has been known that the molecular orbits around $2\alpha$
play important roles.
The molecular orbits in Be isotopes have been suggested in $^9$Be
with $2\alpha+n$ cluster models \cite{OKABE}. 
W. Von Oertzen et al. proposed a picture of Be dimers 
\cite{OERTZEN,OERTZENa}
to understand the excited states of neutron-rich Be isotopes.
They predicted $\sigma$- and $\pi$-orbits which are made from 
linear combinations of $p$-orbits surrounding 2-$\alpha$ clusters 
(see Fig.\ref{fig:sigmapi}).
Itagaki et al. have described the structures of $^{10}$Be and $^{12}$Be
with a extended cluster model by assuming $\alpha$ cores and the 
molecular orbits. 
In the study of Be isotopes
with AMD methods \cite{DOTE,ENYOf,ENYObc}, the structures 
with 2 $\alpha$s and valence neutrons in neutron-rich Be isotopes 
have been microscopically confirmed  without assuming the 
existence of any clusters nor molecular orbits. 
In the present work we investigate 
the behavior of the valence neutrons focusing on the molecular orbits.

\begin{figure}
\noindent
\epsfxsize=0.45\textwidth
\centerline{\epsffile{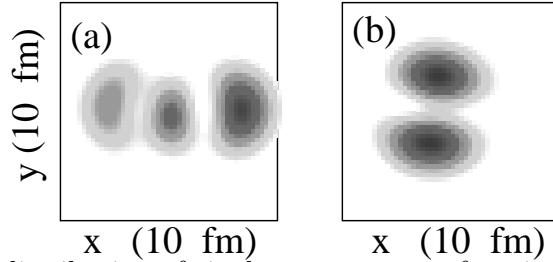}}
\caption{\label{fig:single05p}
The density distribution of single-neutron wave functions
in the intrinsic system of $1/2^+_1$ state 
calculated with case (1). Figure (a) shows the density for the 
highest neutron orbit with 90\% positive parity component, 
while figure (b) is for the second neutron orbit which contains
 90\% of a negative parity component.
}
\end{figure}

By analyzing the ratio of positive and negative parity components
in each single-particle wave function, it is found that all the neutron 
orbits are approximately parity-eigen states, which roughly 
correspond to $1s$-, $1p$-, and $2s1d$ orbits.
The lowest normal-parity $1/2^-$ state comes from a $0\hbar\omega$ 
configuration because 2 neutrons occupy $s$-like orbits and the
other 5 neutrons are in $p$-like orbits.
On the other hand, in the ground $1/2^+$ state with the abnormal parity, 
the last neutron occupies a $sd$-like orbit, which means that the
$1/2^+$ state is dominated by a $1\hbar\omega$ configuration. 
An interesting feature is found in the highest neutron wave function 
which well corresponds to the $sd$-like orbit. 
The density distribution of neutrons are presented in 
Fig.\ref{fig:single05p}(a). It has nodes along the longitudinal direction
of 2-$\alpha$ clusters and contains nearly 90\% of a positive-parity 
eigen state.
It well corresponds to the so-called molecular $\sigma$-orbit explained in 
the schematic figure \ref{fig:sigmapi}(a).
On the other hand, the features of 2 neutron orbits energetically 
below the $\sigma$-orbit are shown in \ref{fig:sigmapi}(b).
The dominating negative-parity component of these orbits 
indicates that the orbits correspond to the $p$-like orbits.
As shown in Fig.\ref{fig:single05p}(b), 
the orbits seem to be similar to the molecular $\pi$-orbits.

\begin{figure}
\noindent
\epsfxsize=0.35\textwidth
\centerline{\epsffile{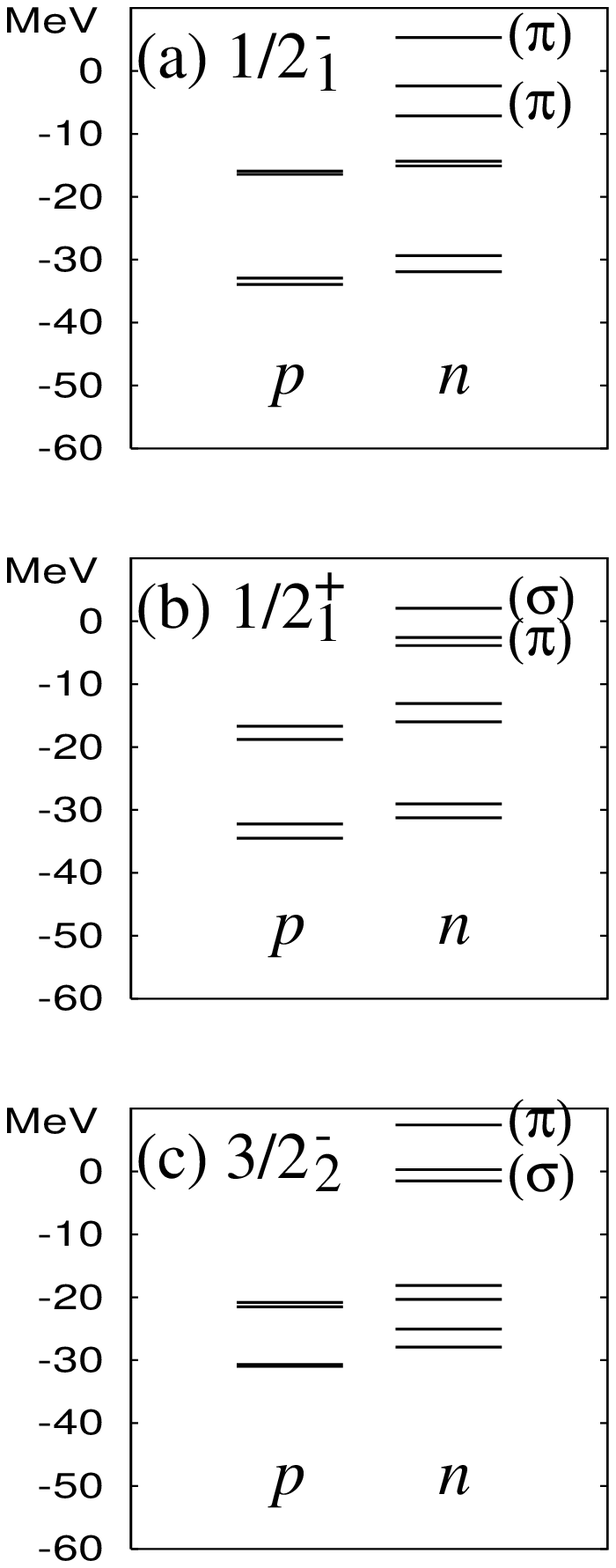}}
\caption{\label{fig:be11hfe}
Single-particle energies in the intrinsic system of 
(a)$1/2^-_1$, (b)$1/2^+_1$, (c)$3/2^-_2$ with the case (1) interaction.
Energies for protons(neutrons) are presented in left(right)
in each figure.
}
\end{figure}

One of the reasons for parity inversion of $^{11}$Be is considered to 
be because of the molecular $\sigma$-orbit which originates from $sd$-orbits.
Since Be isotopes prefer to 
prolate deformations because of 2-$\alpha$ clusters,
the $\sigma$-orbit can easily gain its kinetic energy in the 
clustering developed system. In another word, 
the 2$\alpha$-clustering structure
 is one of advantages for the $\sigma$-orbit in Be isotopes.
From this point of view, it is easy to understand the reason for 
the clustering development in the $1/2^+_1$ state 
in terms of the energy gain of the $\sigma$-orbit in a clustering system.
 Figures \ref{fig:be11hfe}(a),(b),(c) show the single-particle energies 
in the intrinsic states of $1/2^-_1$, $1/2^+_1$ and $3/2^-_2$ 
with the case (1) interaction, respectively. 
 The highest orbit in Fig.\ref{fig:be11hfe}(b)
corresponds to the $\sigma$-orbit in $1/2^+_1$ state. The energy of the 
intruder $\sigma$-orbit is lower than the
highest $\pi$-orbit in $1/2^-_1$.
In all the states:$1/2^-_1$, $1/2^+_1$ and $3/2^-_2$,
the 2-$\alpha$ cores consist of 4 neutrons and 4 protons in 
the $1s$-like orbits and the lower 
$p$-like orbits  seen in Fig.\ref{fig:be11hfe}.
The behavior of the single-particle wave functions of valence neutrons
in the present results is almost consistent with the previous AMD studies
\cite{DOTE,ENYOdoc}.

One of new discoveries in the present work is that the possible 
negative-parity band $K=3/2^-$ has a largely deformed shape 
with a remarkable clustering structure. With the same analysis 
of the single-particle wave
 functions as mentioned above, we find that 
2 neutrons in the $3/2^-_2$ state occupy so-called $\sigma$-orbits.
In another word, the well-developed clustering is caused by 
these 2 neutrons in the $\sigma$-orbits so as to gain their kinetic energies.
It is surprising that such a 2 $\hbar\omega$ state appears 
in the low energy region.
The clustering structure in the $K=3/2^-$ band 
is the very molecular structure which has been predicted 
by W. Von Oertzen \cite{OERTZENa}. This is the first microscopic calculation
which shows that such an exotic structure should exist 
in the low energy region.

In the single-particle energies shown in Fig.\ref{fig:be11hfe}(c), we find 
another interesting thing that the energies of the $\sigma$-orbits 
in $3/2^-_2$ state become lower than that of an odd neutron $\pi$-orbit.
One of reasons for the higher energy of the odd neutron $\pi$-orbit than 
the $\sigma$-orbits is a lack of the pairing 
interaction because of the absence of the other neutron in the
$\pi$-orbit. 
Another reason is the degeneration between $\sigma$-orbits and $\pi$-orbits
at the large inter-cluster distance.
With the increase of relative distance between 2-$\alpha$ clusters, the 
$\sigma$-orbits gain their energy, while the
$\pi$-orbits loose their potential energies.
As the relative distance increases, the $\sigma$-orbits cross the higher
$\pi$-orbits and come down near the lower $\pi$-orbits.
It is helpful
to see the neutron level scheme of two center shell model
\cite{EISENBERG}
to understand the inversion of $\pi$- and the $\sigma$-orbits 
with the increase of inter-cluster distance.

In the single-particle energies shown in Fig.\ref{fig:be11hfe},
the small level spacing between the lower $1s$-orbits and the 
lower $1p$-orbits indicates
the development of 2 $\alpha$ clusters in $1/2^+$ and $3/2^-_2$ states.
In fact, the level spacing is smallest in $3/2^-_2$ state.

As mentioned above, we see the drastic change of neutron structures 
between three rotational bands. It is found that the origins of drastic 
change are the neutrons in $\sigma$-orbits and $\pi$-orbits 
surrounding 2-$\alpha$ cores.
The intrinsic states are characterized by the number of valence neutrons 
in the $\sigma$-orbit. That is to say that, no neutron, 
one neutron and two neutrons occupy the 
$\sigma$-orbits in the $K=1/2^-$ $K=1/2^+$ $K=3/2^-$ bands, respectively. 
The deformation with clustering structure
is enhanced by the increase of the neutrons which occupy 
the $\sigma$-orbits due to the energy gain of $\sigma$-orbits
in the developed clustering structure.

\subsection{Breaking of $2\alpha$ cluster cores}

Although the structures with 2 $\alpha$-cluster cores are seen 
in most of the states of $^{11}$Be,
the $\alpha$ clusters are not the ideal $\alpha$ clusters 
with simple $(0s)^4$ configurations. 
The $\alpha$ clusters are broken slightly because of the spin-orbit force.
 The components of cluster breaking allows the $\beta$ decays
from $^{11}$Li into $^{11}$Be. This is a similar situation as seen in 
the 3$\alpha$-cluster structures of $^{12}$C to which 
Gamow-Teller transitions from $^{12}$B and $^{12}$N are not weak.

In the calculated results, we can know the degrees of 
breaking from the ideal 2-$\alpha$ clusters by estimating
non-zero total-intrinsic spin($S_p\ne 0$) components for protons.
If we neglect $S_p\ge 2$ components and
assume that only $S_p=0$ and $S_p=1$ 
components are contained in a state, the value 
$0.5\times \langle S^2_p \rangle$ indicates directly the mixing ratio of   
the non-$(0s)^4$ configurations, which signifies the degrees of 
cluster breaking in the state. 
 In table 
\ref{tab:protonspin}, the expectation values of the squared total-intrinsic 
spin of protons are listed. 
Generally speaking, the clustering breaking is larger in 
the results in the case (1) than in the case (2) because the 
stronger spin-orbit forces in the case (1) give larger effects 
on the dissociation of 2 $\alpha$-cluster cores.
The states $1/2^-_1$ ,
$3/2^-_1$ and $5/2^-_1$ in $K=1/2^-$ contain 
significant components of the cluster breaking in the ratios of
7\%, 10\% and 20\%, respectively. 
The mixing of cluster breaking allows the $\beta$ decays into
these states from $^{11}$Li, which consistent with the experimental
measurements (Table.\ref{tab:be11beta}). The smaller mixing in $1/2^-_1$ than 
the other $3/2^-_1$ , $5/2^-_1$ states is reflected on the relatively 
weak $\beta$ decays to $1/2^-_1$ as mentioned in the previous section 
on the $\beta$-decay strength.
On the other hand, the breaking of $\alpha$ clusters are very small 
in $K=1/2^+$ and $K=3/2^-$ which have the well-developed clustering 
structures. Therefore Gamow-Teller transitions to the excited states in 
$K=3/2^-$ band from $^{11}$Li are predicted to be weak  except for   
the $\beta$ decays to the $3/2^-_2$ state because this state 
slightly contains
the cluster breaking due to the state mixing with the lower 
$3/2^-_1$ state. 

\begin{figure}
\noindent
\epsfxsize=0.45\textwidth
\centerline{\epsffile{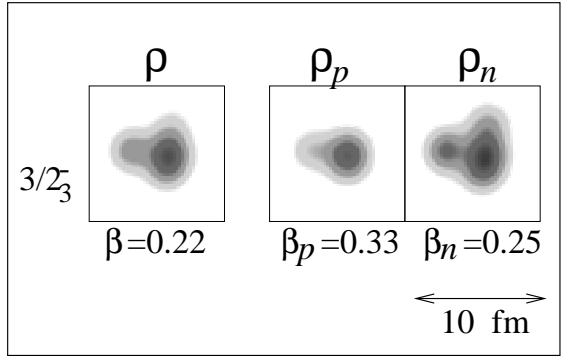}}
\caption{\label{fig:be11-l3n}
The density distribution of the intrinsic wave function of 
$3/2^-_3$ calculated with the case(1) interaction. }
\end{figure}

We discover an excited $3/2^-_3$ state where the $2\alpha$-cluster structure
is completely broken. 
In the density distribution of protons we can not recognize 
the dumbbell shape in this state 
(see Fig.\ref{fig:be11-l3n}).
As seen in Table \ref{tab:protonspin}, the 
main component is the proton intrinsic spin $S_p=1$ state.
As a result, the $\beta$-decay transitions from $^{11}$Li to 
the  $3/2^-_3$ state are 
strong comparing with the $\beta$-decay strength to the other excited states 
of $^{11}$Be. It is consistent with 
the strong $\beta$ decays
 to the state at 8.04 MeV, which have been  recently measured.

Although the 2$\alpha$-core structures are dominant in most of 
Be isotopes, it is interesting that the excited state with 
the completely breaking of one of 2 $\alpha$s appears in neutron-rich Be.
Even in the excited states with 2-$\alpha$ cores, the components of 
the cluster breaking play an important role in the strength of $\beta$ decays 
from $^{11}$Li.

\begin{table}
\caption{\label{tab:protonspin} The expectation values of the squared 
total intrinsic spin of protons $S^2_p$ in the excited states of $^{11}$Be.
}

\begin{center}
\begin{tabular}{lccc}
 $J^\pm_n$ states & $0.5\times \langle S^2_p \rangle$ & \\
         & case(1) & case(2) &\\
\hline
($K=1/2^-$ band) & \\
$1/2^-_1$  &  0.07 & 0.04\\
$3/2^-_1$  &  0.1  & 0.04\\
$5/2^-_1$  &  0.2 & 0.04 \\
\hline
($K=3/2^-$ band) & \\
$3/2^-_2$  &  0.04 & 0.003\\
$5/2^-_2$  &  0.01 & 0.003\\
$7/2^-_2$  &  0.01 & 0.002\\
\hline
($K=1/2^+$ band) & \\
$1/2^+_1$  &  0.03 & 0.01 \\
$3/2^+_1$  &  0.02 & 0.01 \\
$5/2^+_1$  &  0.03 & 0.01 \\
$7/2^+_1$  &  0.02 & 0.01 \\
$9/2^+_1$  &  0.02 & 0.004\\
\hline
$3/2^-_3$  & 0.7 & 1.0 \\
\end{tabular}
\end{center}
\end{table}

\subsection{parity inversion}
We consider details of parity inversion of the ground
state in the case(1) calculations.
Concerning the energy gain of the positive parity $1/2^+$ state,
the effects of the halo structure, the clustering structure with a
prolate deformation, the three-body force, the angular-momentum 
projection, and the paring
have been discussed in the pioneering works \cite{DOTEb}. 

In the present calculations the effects of clustering structure are 
included as well as the pioneering AMD studies \cite{DOTEb,ENYObc}. 
One of the reasons for parity inversion is the energy gain of 
the molecular $\sigma$-orbit in the system with 
developed 2$\alpha$ clustering. 
The details of clustering and molecular orbits 
have been already  mentioned above.

The three-body force was calculated approximately with 
a perturbative treatment
in the pioneering study\cite{DOTEb}. In the present work, 
the three-body force is treated exactly  
as same way as the previous AMD study in Ref.\cite{ENYObc}.

In the previous AMD studies of $^{11}$Be \cite{DOTEb,ENYObc}, 
the angular-momentum projection was perturbatively treated. That is to say
that the energy variation was performed after a parity projection but 
before an angular-momentum projection(VBP).
In the present results, we perform energy variation exactly for the
spin-parity projected wave functions(VAP). Then we find that the 
effect of angular-momentum projection on the parity inversion is 
quite different from the previous VBP results with 
perturbative treatments. 
In order to quantitatively discuss the effects, 
we estimate the energy difference between 
$1/2^-$ and $1/2^+$ states: $\delta E\equiv E_{1/2^+}-E_{1/2^-}$. 
The parity inversion is described by a negative value of $\delta E$.
The experimental value of $\delta E$ has been known to be $-0.32$ MeV.
For simplicity we do parity-projected AMD calculations 
with fixed intrinsic spins
by using the case(1) interaction.
In table \ref{tab:be11vbp}, the values of $\delta E$ and energies 
(a) without angular-momentum projections, (b) the angular-momentum 
projections after variation within the perturbative treatments of VBP, 
and (c) results of the exact 
VAP calculations are listed. Before the angular-momentum projections, 
$\delta E$ is 0.6 MeV. After the projection within 
the perturbative treatments of VBP, $\delta E$ becomes small 
as $-1.1$ MeV. It is because that 
the energy gain of the positive-parity state by the angular-momentum projection
is 1.7 MeV larger than the gain of the negative-parity state. 
The origin is the 
well-developed clustering structure in the former state comparing
with the latter state because a deformed system gains much energy 
due to a angular-momentum projection in general. 
 The VBP results are consistent with the previous 
AMD studies in Refs. \cite{ENYObc,DOTEb}. However when we 
exactly treat the angular-momentum projection with VAP calculations we find
a new aspect in the effects on the energy gain. 
In the previous AMD studies, it is known 
that in many cases VAP results of low-lying states 
is not so much different from VBP results.
It is true in the $1/2^+$ state. However it is not true in the $1/2^-$ state.
In the VAP calculations,
the negative parity state gains much energy as 8.6 MeV due to the 
angular-momentum
projection because the deformation of $1/2^-$ state grows more largely in the 
VAP calculations than in the VBP calculations 
so as to gain the energy of the angular-momentum projected state. 
As a result, $\delta E$ of (c) is 1.0 MeV as much as the 
result before projections (a). 
In fact, the clustering of $^{11}$Be($1/2^-_1$)
is larger in the VAP results than as is expected in the VBP calculations.
Therefore the effect of the angular-momentum projection is not positive 
for the parity inversion in $^{11}$Be.   
This is one of new things in the present work which points that 
the perturbative treatments of angular-momentum projections 
do not always work well for precise descriptions.

\begin{table}
\caption{\label{tab:be11vbp} Energies of the positive and negative parity
states of $^{11}$Be. The parity projected AMD calculations
with fixed intrinsic spins 
(a) before angular-momentum projections, (b) after angular-momentum
 projections of VBP within the perturbative treatment, and (c) 
results of the exact VAP calculations
are listed. We also show (d) the present VAP calculations with
variational intrinsic spins but no diagonalization of the basis. 
The adopted interactions are the case(1) force.
The energy difference $\delta E$ between the positive and negative parity 
states is defined as $E_{1/2^+}-E_{1/2^-}$ for the calculations after
angular-momentum projections and $E(+)-E(-)$ before the projection.}
\begin{center}
\begin{tabular}{ccccc}
& (a) before projection & (b) VBP & (c)VAP & (d)VAP\\
intrinsic spins &  fixed &fixed &fixed &variational parameters\\
\hline
& & & &  \\
 $^{11}$Be(+) &  $-46.5$  & $-53.1$  & $-54.7$  &  $-57.1$ \\
 $^{11}$Be($-$) &  $-47.1$  & $-52.0$  & $-55.7$  &  $-57.0$  \\
$\delta E$ & 0.6 & $-1.1$ & 1.0 & $-0.1$ \\
\end{tabular}
\end{center}
\end{table}

Comparing with the previous AMD studies of low-lying states of $^{11}$Be
\cite{DOTEb,ENYObc},
another improvement in the present framework is treating the intrinsic 
spin functions as variational parameters.
In the Table\ref{tab:be11vbp},
the effect of the flexible intrinsic spins are seen in the difference between
(c) the VAP calculations with the fixed intrinsic spins and 
(d) the VAP calculations with the variational intrinsic spins.
The treatment of flexible intrinsic spins gives an important effect
on the parity inversion of $^{11}$Be. It reduces $\delta E$
to a 1.1 MeV smaller value in (d) than that in the calculations (c).
The importance of the flexible intrinsic spins in 
$p_{3/2}$ shell-closure states 
has been already argued in the studies of $^{12}$C 
\cite{ENYOe} by one of the authors. In the case of $^{11}$Be,
 it is natural to 
consider that the $p_{3/2}$ sub-shell effect is one of the reasons for
the parity inversion because the $p_{1/2}$-orbit sould be raised 
relatively higher than $2s_{1/2}$-orbits.
Unfortunately, the simple AMD method with the fixed intrinsic
spins describe the sub-shell effects insufficiently.
This is one of the advantages of the present framework which can 
describe well both aspects of the clustering and shell effects.

Regarding the other reasons of parity inversion, Sagawa et al. \cite{SAGAWAb}
and Dot\'{e} et al. \cite{DOTEb} suggested the paring effects 
in the neutron $p$-shell. In the present framework, a part of the 
paring effects in $p$-shell should be included automatically by the 
spin-parity projections and the superpositions.

As mentioned above,
the clustering effects, the three-body forces, the 
$p_{3/2}$ sub-shell effects are included  in the present calculations. 
The paring effects are expected to be partially contained.  
The other effect which should be important for parity inversion is  
the neutron-halo effect\cite{SAGAWA,DOTEb}. 
According to the study in Ref.\cite{DOTEb}, 
the effect due to the neutron halo structure is estimated as 0.6 MeV 
reduction of $\delta E$. However the halo effect is
 not taken into account in the present calculations. 
We think that this halo effect and the 
residual paring effect are contained effectively in the strength 
parameter $u_{LS}=3700$ MeV of the spin-orbit force which is  
slightly stronger that the parameter $u_{LS}=3000$ MeV adopted in 
Ref.\cite{ENYOe}. 
The halo structures should be important only for the loosely bound
states with low-spin orbits such as $s$-orbits and $p$-orbits at most.
The present strong effective spin-orbit force in case(1) 
may artificially reduce too much the theoretical excitation energies 
of the other states in $K=1/2^+$ band except for the $1/2^+$ state.
Before concluding the relative position of the energies 
between $K=1/2^-$ and $K=1/2+$ 
bands we should more detailed researches 
by taking the remained effects such as the halo structure into 
consideration.

\section{Summary}
 \label{sec:summary}
We studied the structures of the ground and excited states 
of $^{11}$Be with VAP calculations in the framework of antisymmetrized 
molecular dynamics.
Various kinds of excited states with the clustering structures and also 
the non-clustering structures were discovered in the 
theoretical results.
We predicted many excited states. Most of the states belong to
 three rotational bands: 
$K=1/2^+$, $K=1/2^-$ and $K=3/2^-$, which are dominated by $1\hbar\omega$,
$0\hbar\omega$ and $2\hbar\omega$ configurations, respectively. 
It should be pointed out that the formation of 2 $\alpha$-cluster cores
is seen in many excited states of $^{11}$Be in the the present
resutls in spite of no assumption of the existence of clusters.
The interesting point is that an eccentric
rotational band $K=3/2^-$ with the mostly developed clustering 
structure starts from the $3/2^-_2$ state at about 4 MeV and reaches 
high spin states. 

The experimental data concerning the $\beta^+$-decay and $\beta^-$-decay 
strength were 
reproduced well. We also argued that the cluster breaking 
plays an important role to allow the $\beta$ decays from $^{11}$Li. 
The significant breaking of 2-$\alpha$ cores in the states in $K=1/2^-$ band
has been seen in quantitatively estimating the breaking of clusters.
We discovered a non-clustering state at about 10 MeV.
One of the characteristics of this state is the strong $\beta$ decays from 
$^{11}$Li, which well corresponds to the new excited states at 8.04 MeV
found in the $\beta^-$-decay measurements. 

By analyzing the single-particle wave functions in the intrinsic states,
it was found that the molecular $\sigma$-orbits 
surrounding 2-$\alpha$ cores play important roles in the clustering 
structures of $^{11}$Be.  
In the ground band $K=1/2^+$, one neutron
occupies the $\sigma$-orbits. In the newly predicted $K=3/2^-$ band 
is dominated by $2\hbar\omega$ configurations with two neutrons in the
$\sigma$-orbit. When the surrounding neutrons occupy the $\sigma$-orbits, 
the clustering development is enhanced so as to gain their 
kinetic energy.
In another word, one of the reasons for the parity inversion and the 
low-lying $2\hbar\omega$ states is the energy gain of the $\sigma$-orbits
with the developed clustering structures.

Concerning the mechanism of the parity inversion,
we described the importance of molecular neutron orbits 
in the developed clustering structure and also mentioned about
the $p_{3/2}$ sub-shell effect. 
Although the spin parity of the ground state is described with 
a set of interaction parameters case (1), we need more detailed researches
taking the halo structures of the ground state into account. 

\acknowledgments
The authors would like to thank Dr. N. Itagaki and Dr. Dot\'{e} 
for many discussions.
They are also thankful to Professor W. Von Oertzen for helpful discussions 
and comments. 
The computational calculations of this work are supported by 
the Supercomputer Project No.58, No.70 of High Energy Accelerator Research
Organization(KEK), and Research Center for Nuclear Physics 
in Osaka University.


\end{document}